\documentclass[prd,twocolumn,showpacs,floatfix,amsmath,nofootinbib,amssymb,floatfix]{revtex4-1}
\usepackage{graphicx,color,dcolumn,booktabs,bm}
\usepackage{longtable}
\usepackage{txfonts}
\usepackage{overpic}
\usepackage{amssymb}
\usepackage{indentfirst}
\usepackage{feynmf}   
\usepackage{slashed}  
\usepackage{cases}
\usepackage{color}
\usepackage{multirow}
\usepackage{epstopdf}
\usepackage{graphicx,color,dcolumn,booktabs,bm}
\usepackage[colorlinks,
            citecolor=blue,
            anchorcolor=green,
            menucolor=orange,
            linkcolor=red,
            filecolor=red,
            runcolor=pink,
            urlcolor=blue,
            frenchlinks=red]{hyperref}

\graphicspath{{Figures/}} %
\def\half{{\textstyle{1\over 2}}}
\def\thalf{{\textstyle{3\over 2}}}
\def\fhalf{{\textstyle{5\over 2}}}
\allowdisplaybreaks

\begin{document}

\title{Understanding spin parity of $P_c(4450)$ and $Y(4274)$ in a hadronic molecular state picture}
\author{Jun He}
\email{junhe@impcas.ac.cn}
\affiliation{Department of  Physics and Institute of Theoretical Physics, Nanjing Normal University,
Nanjing, Jiangsu 210097, China}
\affiliation{Nuclear Theory Group, Institute of Modern Physics, Chinese Academy of Sciences, Lanzhou 730000, China}

\affiliation{Research Center for Hadron and CSR Physics, Lanzhou University
and Institute of Modern Physics of CAS, Lanzhou 730000, China}

\begin{abstract}

The hidden-charmed pentaquark $P_c(4450)$ and the charmonium-like
state $Y(4274)$ are investigated as a $\bar{D}^*\Sigma_c$  and a
$D_s\bar{D}_{s0}(2317)$  molecular state, respectively. The spin
parities of these two states cannot be well understood if only S-wave
$\bar{D}^*\Sigma_c$ and  $D_s\bar{D}_{s0}(2317)$ interactions are
considered. In this work, the  interactions are studied in a
quasipotential Bethe-Salpeter equation approach with a partial wave
decomposition on spin parity $J^P$, and the contributions of different
partial waves are studied in a two-channel scattering model including
a generating channel and an observation channel. Two poles at
$4447\pm4i$ and $4392\pm46i$ MeV are produced  from the
$\bar{D}^*\Sigma_c$ interaction coupled with the $J/\psi p$ channel in
$3/2^-$ wave and $5/2^+$ wave, respectively. The peak for the $5/2^+$
state has a comparable height as that of the $3/2^-$ state in the $J/\psi
p$ invariant mass spectrum. The $D_s\bar{D}_{s0}(2317)$ interaction
coupled with the $J/\psi\phi$ channel is studied  and a pole at
$4275\pm11i$ MeV is produced in $J^{P}=1^{+}$ wave, which corresponds
to P-wave $D_s\bar{D}_{s0}(2317)$ interaction. The pole from S-wave
$D_s\bar{D}_{s0}(2317)$ interaction is far below that from P-wave
interaction even the $J/\psi\phi$ threshold, so cannot be observed in
the $J/\psi\phi$ channel.  The result suggests that in these cases
a state carrying a spin parity corresponding to P-wave interaction
should be taken as seriously  as these carrying a spin parity
corresponding to S-wave interaction in the hadronic molecular state
picture.
\end{abstract}

\pacs{14.20.Pt, 03.65.Nk, 11.10.St}
\keywords{}

\maketitle


\section{Introduction}\label{sec1}

The hadronic molecular state picture is one of the most popular
interpretations of the exotic state in market~\cite{Chen:2016qju}. It
has been widely applied to explain a series of experimentally observed
exotic states, which cannot be assigned in the conventional quark model
but is close to the threshold of two hadrons. In the literature,
people often focus on the bound state from S-wave interaction and
assume the P-wave bound state should be difficult to form from
hadron-hadron interaction and to observe in experiment.  For example,
the $X(3872)$ and the $Z_c(3900)$ are related to  isoscalar and
isovector S-wave $D\bar{D}^*$
states~\cite{Tornqvist:2004qy,Close:2003sg,He:2015mja}, and the
$Y(4274)$ and the $Y(4140)$ are related to S-wave
$D_s\bar{D}_{s0}(2317)$ and $D_s^{*+}D_s^{*-}$ states,
respectively~\cite{Liu:2010hf,He:2011ed,He:2013oma,Liu:2009ei}. There
also exist predictions of  hidden-charmed pentaquark from S-wave
anticharmed meson and charmed baryon
interactions~\cite{Wu:2010jy,Yang:2011wz}.

The recent observation of the  $P_c(4450)$ and $P_c(4380)$ at LHCb
confirmed the existence of the hidden-charmed pentaquark. With the
help of partial wave analysis, LHCb provides the information about
spin parities as well as masses of these
states~\cite{Aaij:2015tga,Aaij:2016iza,Aaij:2016nsc}. Surprisingly,
different from the predictions in Refs.~\cite{Wu:2010jy,Yang:2011wz}
the hidden-charmed pentaquarks $P_c(4380)$ and $P_c(4450)$ carry
opposite parities. It is difficult to explain both states as hadronic
molecular states from relevant S-wave anticharmed meson and charmed
baryon interactions, i.e.,  $\bar{D}\Sigma_c^*$, $\bar{D}^*\Sigma_c$,
and $\bar{D}^*\Sigma^*_c$ interactions. In
Ref.~\cite{Meissner:2015mza}, the authors proposed that the $P_c(4450)$
can be reproduced from S-wave interaction of a proton and a P-wave
charmonium $\chi_{c1}$, with a small coupling to the $J/\psi p$ channel.
It is an interesting interpretation but a little different from the
standard molecular state picture because the attraction is from
transition between the $\chi_{c1}p$ and $J/\psi p$ channel instead of
the direct $\chi_{c1}p$ interaction which is suppressed according to
the OZI rule.  In Ref.~\cite{He:2015cea}, the $\bar{D}\Sigma_c^*$,
$\bar{D}^*\Sigma_c$, and $\bar{D}^*\Sigma^*_c$ interactions were
investigated in a quasipotential Bethe-Salpeter equation approach with
a partial wave decomposition based on $J^P$. In such an approach, the
orbital angular momentum $L$ is not considered explicitly because it
is not a good quantum number when the calculation is relativistic and
the experimental result is provided with spin parity $J^P$ directly.
A bound state in $5/2^+$ wave  is produced from  the
$\bar{D}^*\Sigma_c$ interaction, which can be explained as
experimentally observed $P_c(4450)$~\cite{He:2015cea}. Such spin
parity cannot be produced from S-wave interaction of a $\bar{D}^*$
with $1^-$ and a $\Sigma_c$ with $1/2^+$. So this bound state with
$5/2^+$ should be from P- and F-wave $\bar{D}^*\Sigma_c$ interaction.
Such a challenge also happens in the case of the $Y(4274)$. The spin
parity $1^{++}$ determined at LHCb~\cite{Aaij:2016iza,Aaij:2016nsc}
conflicts with  previous S-wave $D_s\bar{D}_{s0}(2317)$ molecular
state interpretation~\cite{Liu:2010hf,He:2013oma}, which suggests
P-wave interaction should be also introduced in this case.

In this work we will study the $P_c(4450)$ and $Y(4274)$  in a
quasipotential Bethe-Saltpeter equation with a partial wave
decomposition on spin parity $J^P$. The spin parity which corresponds
to P wave will be considered and focused.  In the literature, there
are some studies about the  P-wave molecular
state~\cite{He:2015cea,He:2010zq,Kang:2016zmv}, especially Ref.
~\cite{Kang:2016zmv} where S-wave interaction is forbidden.  However,
explicit comparison of the effects of P wave and S wave on
experimental observables, such as cross section and invariant mass
spectrum, is scarce.  Hence, in this work, we will focus on three
questions:
\begin{itemize}
  \item Admittedly, P-wave interaction should be weaker than S-wave interaction. We will study whether P-wave interaction is too weak to form a bound state, or  weak but still enough to form a bound state in some cases.
  \item Can the P-wave bound state, if it can be produced,  be observed as these from S-wave interaction?
  \item If the observed state corresponds to the  P-wave bound state,
	  it should be answered where is the S-wave bound state which should be easier to produce.
\end{itemize}

In the next section, the formalism adopted in the quasipotential
Bethe-Salpeter equation approach is presented, and a toy model of
two-channel scattering of scalar mesons is adopted to compare  P-wave
and S-wave contributions. In Sec.~\ref{sec3}, the LHCb pentaquarks
$P_c(4450)$ and $P_c(4380)$ are studied as the $\bar{D}^*\Sigma_c$
molecular states. The $D_sD_{s0}(2317)$  interaction and the $Y(4274)$
are studied in Sec.~\ref{sec4}.  The discussion and summary are given in the last section.

\section{FORMALISM}\label{sec2}

In this work, we will introduce a two-channel scattering, which
includes a generating channel and an observation channel, to study
the relative magnitude of contributions of different spin-parity partial waves.
\begin{itemize}
\item Generating channel: It has a higher threshold and is adopted to
	generate the bound state by exchange of light mesons. In the cases
	considered in this work, the  $\bar{D}^*\Sigma_c$  channel and
	the $D_s\bar{D}_{s0}(2317)$ channel are considered for the
	$P_c(4450)$ and the $Y(4274)$, respectively.
\item  Observation channel: It has a lower threshold and is adopted to
	observe the bound state generated by the generating channel.
	For the observation channels considered in this work, i.e. the
	$J/\psi p$ channel and the $J/\psi \phi$ channel, the
	interaction is very weak according to the OZI rule.
\end{itemize}
The generating and  observation channels are coupled by exchanges
of heavy mesons, $D$ or $D^*$ mesons here. With the transition between
two channels, the bound state generated by the generating channel will
leave the real axis in the complex plane and exhibit itself as a peak in
the invariant mass spectrum of the observation channel. The
contributions of different spin-parity partial waves in the
observation channel can be compared.  To study the  two-channel
scattering, a coupled-channel quasipotential Bethe-Salpeter equation
approach will be adopted.

\subsection{Quasipotential Bethe-Salpeter equation}\label{sec21}

The general form of the Bethe-Salpeter equation for the scattering amplitude reads
\begin{align}
&{\cal M}^{mn}(k'_1k'_2,k_1k_2;P)\nonumber\\&={\cal
V}^{mn}(k'_1k'_2,k_1k_2;P)+\sum_l\int\frac{d^4
k''}{(2\pi)^4}\nonumber\\
&\cdot
{\cal
V}^{ml}(k'_1k'_2,k''_1k''_2;P)G^l(k''_1k''_2){\cal
M}^{ln}(k''_1k''_2,k_1k_2;P), \label{Eq: BS}
\end{align}
where ${\cal V}$ is the potential kernel and $G$ is the product of the propagators
for two constituent particles. Here the momentum of the system $P=k_1+k_2=k'_1+k'_2$,
and the relative momentum $k''=(k''_2-k''_1)/2$. The superscript $l$, $m$ or $n$ remarks
the different channels,  generating and  observation channel here.

The Bethe-Salpeter equation is usually reduced to a three-dimensional
equation with a quasipotential approximation. To study the behavior of
the one-boson-exchange interaction below  threshold, the
off-shellness  of two constituent hadrons should be kept.  Here we
adopt a most economic method, that is, the covariant spectator
theory~\cite{Gross:1991pm,VanOrden:1995eg},  which was explained
explicitly in the appendices of Ref.~\cite{He:2015mja} and applied to
studied the $\Lambda(1405)$, the $Z_c(4430)$, the $N(1875)$, and the
$Z(3900)$ and the LHCb
pentaquarks~\cite{He:2015mja,He:2015cea,He:2014nya,He:2015cca,He:2014nxa,He:2015yva}.
Written down in the center-of-mass frame where $P=(W,{\bm 0})$, the
propagator  is
\begin{align}
	G&=2\pi i\frac{\delta^+(k_2^{~2}-m_2^{2})}{k_1^{~2}-m_1^{2}}
	\nonumber\\&=2\pi
	i\frac{\delta^+(k^{0}_2-E_2({\bm p}))}{2E_2({\bm p})[(W-E_2({\bm
p}))^2-E_1^{2}({\bm p})]},
\end{align}
where $k_1=(k_1^{0},-\bm
p)=(W-E_2({\bm p}),-\bm p)$ and $k_2=(k_2^{0},\bm
p)=(E_2({\bm p}),\bm p)$ with $E_{1,2}({\bm p})=\sqrt{
M_{1,2}^{~2}+|\bm p|^2}$. A definition $G_0=G/(2\pi i)$ will be used
for convenience thereafter. The constituent particle 2, which is the
heavier one,  is put on shell   to satisfy
the charge-conjugation invariance because
the meson-exchange model is adopted in the current work~\cite{Gross:1999pd}. A numerical
discussion about different choices of the oneshell constituent particle
was  made in Ref.~\cite{He:2017aps}, and no obvious differences
were found with different choices.
We would like to note that the covariance and the unitary are kept in this quasipotential approximation.

After multiplying
the  polarized vector and spinor on both sides of  Eq.~(\ref{Eq: BS}),
 we obtain an equation for helicity amplitude as
\begin{align}
i{\cal M}^{mn}_{\lambda',\lambda}({\bm p}',{\bm p})&=i{\cal
V}^{mn}_{\lambda'\lambda}({\bm p}',{\bm p})+\sum_{l,\lambda''}\int\frac{d^3
{\bm p}''}{(2\pi)^3}\nonumber\\
&\cdot i{\cal
V}^{ml}_{\lambda'\lambda''}({\bm p}',{\bm p}'')G^l_0({\bm p}'')~i{\cal
M}^{ln}_{\lambda''\lambda}({\bm p}'',{\bm p}),
\end{align}
where ${\bm p}$, ${\bm p}'$ and ${\bm p}''$ are the momenta of constituent 2. Here and hereafter,  individual helicities are omitted where redundant and  states are only labeled
by the total helicities $\lambda$, $\lambda'$ and $\lambda''$.

In this work, we make a partial wave decomposition of the helicity amplitude ${\cal M}$ based on spin parity $J$ as~\cite{Chung}
\begin{align}
{\cal M}_{\lambda'\lambda}({\bm p}',{\bm p})&=
	\sum
	 _{J\lambda_R}{\frac{2J+1}{4\pi}}D^{J*}_{\lambda_R,\lambda'}(\phi',\theta',0)\nonumber\\
	&\cdot{\cal
	M}^J_{\lambda'\lambda,\lambda_R}({\rm p}',{\rm p})D^{J}_{\lambda_R,\lambda}(\phi,\theta,0),
\end{align}
where $D^{J}_{\lambda_R,\lambda}(\phi,\theta,0)$ is the rotation matrix with $J$ being the angular momentum for the partial wave  considered and $\lambda_R$ being the helicity of the bound state.   A definition ${\rm p}\equiv|{\bm p}|$  is adopted here in order to avoid confusion with the four-momentum $p$.
Without loss of  generality, we choose the scattering to be in the $xz$ plane, the potential is written as
\begin{align}
{\cal V}_{\lambda'\lambda}^J({\rm p},{\rm p}')=2\pi\int d\cos\theta d^{J}_{\lambda\lambda'}(\theta_{p',p})
{\cal V}_{\lambda'\lambda}({\bm p}',{\bm p}),\label{VV}
\end{align}
where the momenta $k_1=(W-E,0,0,-{\rm p})$,
$k_2=(E,0,0,{\rm p})$  and $k'_1=(W-E',-{\rm p}'\sin\theta,0,-{\rm p}'\cos\theta)$,
$k'_2=(E',{\rm p}'\sin\theta,0,{\rm p}'\cos\theta)$.

Besides the above partial wave decomposition, the amplitude with fixed party is introduced as
${\cal M}^{J^P}_{\lambda'\lambda}={\cal M}^{J}_{\lambda'\lambda}+{\eta} {\cal M}^J_{\lambda'-\lambda}$
with $\eta=PP_1P_2(-1)^{J-J_1-J_2}$, where $P$ and $P_{1,2}$ are the parities and $J$ and $J_{1,2}$ are the angular momenta for the system and particle 1 or 2 ~\cite{Penner}.  The partial wave Bethe-Salpeter equation with fixed
spin parity $J^P$ reads as~\cite{He:2015mja}
\begin{align}
i\hat{\cal M}_{\lambda'\lambda}^{mn,J^P}({\rm p}',{\rm p})
&=i\hat{\cal V}^{mn,J^P}_{\lambda'\lambda}({\rm p}',{\rm
p})+\sum_{l,\lambda''}\int\frac{{\rm
p}''^2d{\rm p}''}{(2\pi)^3}\nonumber\\
&\cdot
i\hat{\cal V}^{ml,J^P}_{\lambda'\lambda''}({\rm p}',{\rm p}'')
G^{l}({\rm p}'')~i\hat{\cal M}^{ln,J^P}_{\lambda''\lambda}({\rm p}'',{\rm
p}),\label{Eq: BS_PWA}
\end{align}
where $\lambda$, $\lambda'$ and $\lambda''\geq0$ and $\hat{M}^{J^P}_{\lambda'\lambda}=f_{\lambda'}f_{\lambda} M^{J^P}_{\lambda'\lambda}$, with $f_0=\frac{1}{\sqrt{2}}$ and $f_{\lambda\neq0}=1$.
The potential with fixed parity is of a form
\begin{align}
\hat{\cal V}^{J^P}_{\lambda'\lambda}({\rm p}',{\rm
p})&=2\pi f_{\lambda'}f_{\lambda} \int d\cos\theta [d^{J}_{\lambda\lambda'}(\theta_{p',p})
{\cal V}_{\lambda'\lambda}({\bm p}',{\bm p})\nonumber\\
&+\eta d^{J}_{-\lambda\lambda'}(\theta_{p',p})
{\cal V}_{\lambda'-\lambda}({\bm p}',{\bm p})].\label{Eq: BS_PWAV}
\end{align}

By using  normalization of the Wigner $D$ matrix, the integration of the amplitude is
\begin{eqnarray}
\sum_{\lambda'\lambda}\int d\Omega |{\cal M}_{\lambda'\lambda}({\bm p}',{\bm p})|^2=\sum_{J^P,\lambda'\geq0\lambda\geq0}|\hat{\cal M}^{J^P}_{\lambda'\lambda}({\rm p}',{\rm p})|^2. \label{Eq: amplitudes sum}
\end{eqnarray}
Since there is no interference between the contributions from different partial waves,
total cross section or invariant mass spectrum can also be divided into partial-wave cross sections.

To solve the integral equation (\ref{Eq: BS_PWA}), we discretize the momenta ${\rm p}$,
${\rm p}'$, and ${\rm p}''$ by the Gauss quadrature with a weight $w({\rm
p}_i)$ and have ~\cite{He:2015mja}
\begin{eqnarray}
{M}_{ik}
&=&{V}_{ik}+\sum_{j=0}^N{ V}_{ij}G_j{M}_{jk}.\label{Eq: matrix}
\end{eqnarray}
The propagator $G$ is of a form
\begin{eqnarray}
	G_{j>0}&=&\frac{w({\rm p}''_j){\rm p}''^2_j}{(2\pi)^3}G_0({\rm
	p}''_j), \nonumber\\
G_{j=0}&=&-\frac{i{\rm p}''_o}{32\pi^2 W}+\sum_j
\left[\frac{w({\rm p}_j)}{(2\pi)^3}\frac{ {\rm p}''^2_o}
{2W{({\rm p}''^2_j-{\rm p}''^2_o)}}\right],
\end{eqnarray}
with on-shell momentum
\begin{eqnarray}{\rm p}''_o=\frac{1}{2W}\sqrt{[W^2-(M_1+M_2)^2][W^2-(M_1-M_2)^2]}.\label{Eq: mometum onshell}
\end{eqnarray}

In this work, we will search for the  pole of  scattering amplitude as $M=(1-{ V} G)^{-1}V$. By analytic continuation
into complex plane $W\to z$,  the pole can be found by variation of
$z$ in the
complex plane to satisfy $|1-V(z)G(z)|=0$.

\subsection{Toy model: Two-channel scattering of scalar mesons}\label{sec22}

Since realistic interaction is complex, we present first a simple
toy model to explain why we should not work with S-wave interaction
only. Here, the generating channel is composed of two scalar particles
with equivalent masses $M=2.2$ GeV and observation channel is composed
of two scalar particles with masses 1 and 3 GeV. All particles
involved are assumed to be scalar and isoscalar particles for simplicity.
In the current case, the $0^+$ and $1^-$ waves with partial wave
decomposition on spin parity $J^P$ correspond to S and P waves,
respectively.

The potential in the one-boson-exchange model for the two-channel
scattering is written as,
\begin{equation}
i{\cal V}=\left(\begin{matrix} \dfrac{C}{q^2-m^2}&\dfrac{C'}{q^2-m'^2}\\\dfrac{C'}{q^2-m'^2}&0
\end{matrix}\right),\label{Eq: toy V}
\end{equation}
where $q$ is
momentum of  exchanged meson. The $C$ and $C'$ describe the strengths of interactions for the
generating channel  and transition of two channels, respectively. In the cases of the $\bar{D}^*\Sigma-J/\psi p$ and
$D_s\bar{D}_{s0}(2317)-J/\psi \phi$  interactions considered in the current
work, the interaction in
the generating channel is mediated by exchanges of  light mesons,  such as
$\pi$ and $\rho$ mesons, and the coupling of generating and
observation channel by exchanges of charmed mesons, such as the $D$  meson.
To connect with these realistic cases,  in the toy model  we choose
$m=0.5$ GeV and $m'=2$ GeV, which are at the same order of  masses of
light meson and of charmed meson, respectively.

Before presenting the explicit numerical results about the S and P
waves, we give a simple analytical discussion.  According to
the definition of partial-wave potential in Eq.~(\ref{VV}), for a scalar system considered here the S-wave
contribution is from the terms with $(\cos\theta)^0=1$, and the P-wave
contribution is from the terms with $\cos\theta$ which usually appears
with a factor  of $\delta^2=({\rm p}/M)^2$. If both constituents are
on shell, we will have a very small $\delta$ because of the small
binding energy of molecular state of an order of
10 MeV and large masses $M$ of two constituent hadrons  of the order
of 1 GeV . It is why the suppression
of the  P-wave contribution seems  obvious in the  hadron physics community.

Now we take the potential of generating channel as an example to show why
such an analysis is not so reliable for meson-exchange potential. If we assume $\delta$  small the
potential can be written as
\begin{eqnarray}
\frac{C}{q^2-m^2}&=&\frac{C}{2M^2-2E_2({\rm p})~E'_2({\rm p}')+2{\rm p}{\rm p}'\cos\theta-m^2}\nonumber\\
&\approx&\frac{-C}{{\rm p}^2+{\rm p}'^2+m^2-2{\rm p}{\rm p}'\cos\theta}.
\end{eqnarray}
If the momentum ${\rm p}^{(')}$ is also much smaller than the mass of the
exchanged meson $m$, the term $2pp'\cos\theta$ can be omitted so that
the P-wave contribution vanishes.  However, the exchanged meson is
often light in the one-boson-exchange model. Furthermore, except at threshold it is
impossible to put both constituent particles on shell. The momentum is
not fixed but a variable of integration in our quasipotential
Bethe-Saltpeter approach and popularly used Lippmann-Schwinger
equation approach. It is often cut off at about 1 GeV, so the
momentum may be larger than the mass of the  exchanged meson.

\begin{figure}[h!]
\begin{center}
\includegraphics[bb= 150 588 430 730,clip, scale=1.03]{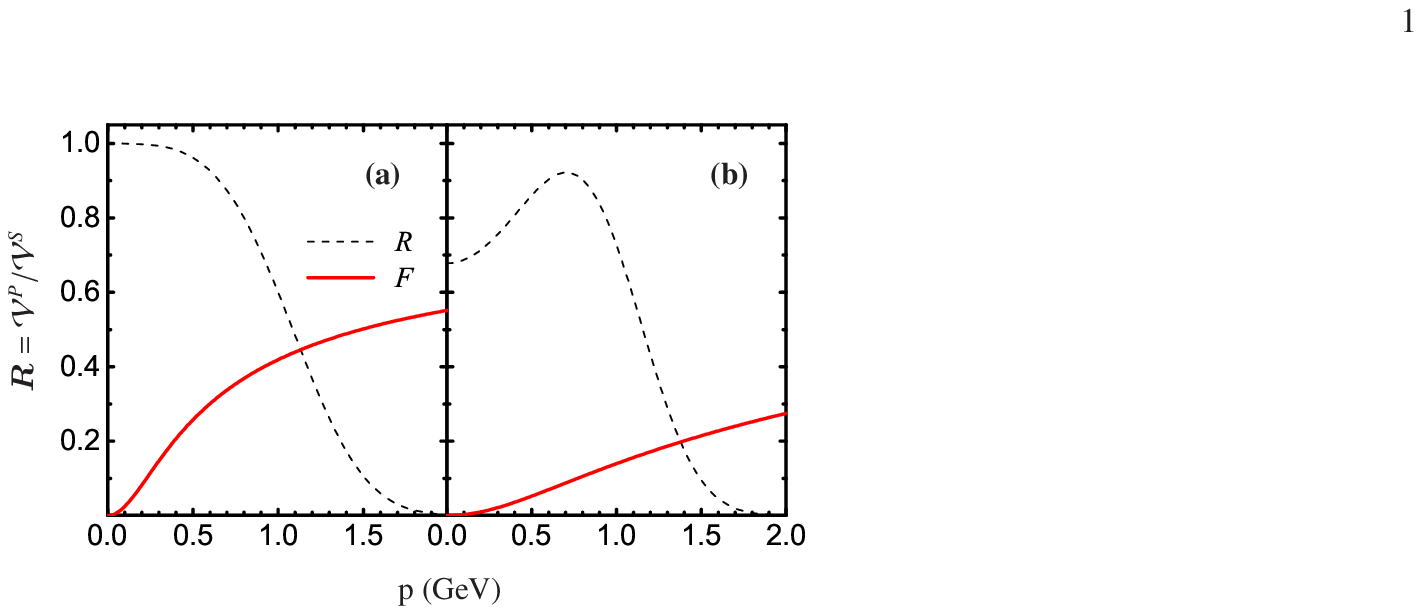}
\caption{(Color online) The ratio $R$ between  P-wave  and S-wave
potential. Part (a) is for  the generating channel and Part (b) for the  transition between the generating  and the observation channel.
\label{Fig: R}}
\end{center}
\end{figure}

In the chiral unitary approach, the cutoff in momentum  can be seen as
regularization~\cite{Oset:1997it}, which can be related to the
dimensional regularization as discussed in Ref.~\cite{Oller:1998hw}.
Inserting the potential in Eq.~(\ref{Eq: toy V}) into the partial-wave
Bethe-Salpeter equation in Eq.~(\ref{Eq: BS_PWA}), one can find that
the convergence is not satisfied in our approach.  Hence, a
regularization is also needed  in our approach.  We will adopt an
exponential regularization by introducing a form factor of exponential
form in the propagator as
\begin{eqnarray}
	G_0({\rm p})\to G_0({\rm p})F({\rm p})=G_0({\rm p})\left[e^{-(k_1^2-m_1^2)^2/\Lambda^4}\right]^2,\label{Eq: FFG}
\end{eqnarray}
with $k_1$ and $m_1$ being the momentum and mass of the charmed
meson~\cite{He:2015mja}. The particle 2 is not involved because of its
on-shell-ness. With such regularization, the momentum usually spreads
from 0 to about 1 GeV as presented  in Fig.~\ref{Fig: R}.

Here a binding energy $E=10$ MeV is adopted and cutoff $\Lambda$ is
chosen at 2 GeV. For convenience, we plot the results with ${\rm
p}={\rm p}'$.  The form factor $F({\rm p})$ is about 50\% at  ${\rm
p}\approx$1 GeV.  So, it is unreliable to regard the P-wave
contribution as a negligible contribution even for  exchange of
$\rho$, $\omega$, $\sigma$ or $\phi$ mesons, which is not so light as
the $\pi$ meson.

In Fig.~\ref{Fig: R}, the ratio $R={\cal V}^{P}/{\cal V}^{S}$ between
the P-wave  and S-wave potential with ${\rm p}={\rm p}'$ is also
depicted.  It is obvious that the P-wave potential is smaller but of
the same order of magnitude as the S-wave potential for exchange of a
particle with mass $m=0.5$ GeV. The ratio $R$ for the transition of
the generating  and  observation channel is smaller than that for
the generating channel, which is due to larger mass of the exchanged meson.

\begin{figure}[h!]
\begin{center}
\includegraphics[bb=180 578 430 740, clip, scale=1]{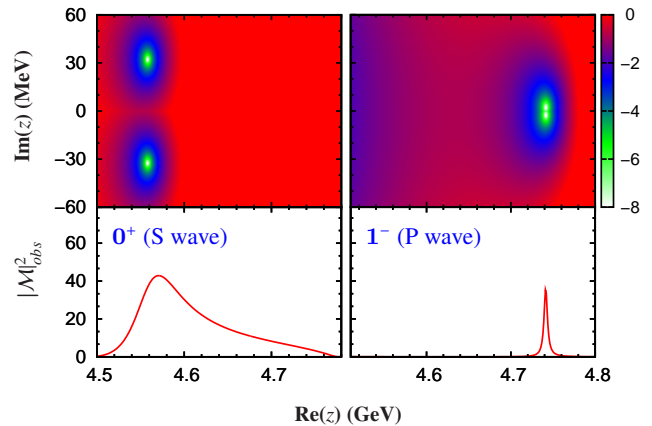}
\caption{The $\log|1-V(z)G(z)|$ and square of scattering amplitude
$|{\cal M}|^2_{obs}$ for the toy model. The results in $0^+$ (S wave) (left panel) and $1^-$ (P wave) (right panel) are drawn to the same scale. \label{Fig: toy model}}
\end{center}
\end{figure}

The above analysis suggests that P-wave interaction is weaker than S-wave
interaction but still promising to produce a bound state.  In the
following, we will make an explicit calculation to compare  P-wave and
S-wave  contribution by taking explicit strengthes $C=6000$ GeV  and
$C'=2000$ GeV as an example. With potential in Eq. (\ref{Eq: toy V}),
the $\log |1-V(z)G(z)|$ is plotted  in Fig.~\ref{Fig: toy model}  with variations of  Re($z$) and Im($z$). The poles can be identified from the plot at $z$ which satisfies $|1-V(z)G(z)|=0$. The square of the scattering amplitude for the observation channel
\begin{align}
|{\cal M}^{J^P}_{obs}|^2=\sum_{\lambda'\geq0\lambda\geq0}|\hat{\cal M}^{J^P}_{obs, \lambda'\lambda}({\rm p}'_o,{\rm p}_o)|^2,\label{Eq: square amplitude}
\end{align}
which is the core of many observables, such as cross section and invariant mass spectrum, is  also presented  in Fig.~\ref{Fig: toy model}.
The scattering amplitudes   $\hat{\cal M}^{J^P}_{obs, \lambda'\lambda}({\rm p}'_o,{\rm p}_o)$ correspond
to the observation-channel part of amplitude $\hat{\cal M}^{J^P}_{\lambda'\lambda}({\rm p}', {\rm p})$  with fixed spin 
parity in Eq.~(\ref{Eq: amplitudes sum}) with the momenta ${\rm p}'$ and ${\rm p}$ being chosen as 
on shell momentum defined in Eq.~(\ref{Eq: mometum onshell}). The scattering amplitude is obtained by a numerical solution of the Bethe-Salpeter equation with  fixed spin parity in Eq.~(\ref{Eq: BS_PWA}) by transforming it to a matrix equation (\ref{Eq: matrix}).

A P-wave bound state is produced from the generation channel as well
as a S-wave state. The P-wave state is closer to the threshold of the
generating channel than the S-wave state, which is easy to understand
because the P-wave interaction is, though strong enough to produce a
bound state, but still weaker than the S-wave interaction. Such
a phenomena will be helpful to answer where is  S-wave bound state which
should be easier to produce if the observed state corresponds to a
P-wave bound state. It will be discussed  later in the realistic case in the next section.

If only the generation channel is considered, the pole for the bound
state is at the real axis. After transition of generating and
observation channels is included, the poles for both S- and P-wave abound states leave the real axis to the complex plane. In other words, the transition will give the bound state width.
In the model considered here,  the S-wave and the P-wave states decay
to the observation channel  through S and P waves, respectively. The P-wave
state has smaller width than the S-wave state because the transition
between generating and observation channel in P wave is weaker than
those in S wave as shown in Fig.~\ref{Fig: R}. However, such weakness
in P wave has a relatively small affect on the height of peaks observed.
The peak of the P-wave state  observed in the observation channel is
almost the same heigh as that of the S-wave state. Based on the results
we concluded that the P-wave interaction is weaker but may be still
enough to form an observable bound state, at least for the toy model considered here.

\section{Application to LHCb hidden-charmed pentaquarks}\label{sec3}

Now, we turn to a realistic case, the LHCb hidden-charmed pentaquarks.
If these two pentaquarks are interpreted as the $\bar{D}\Sigma^*$ or
$\bar{D}^*\Sigma$ molecular state, the opposite parities suggest that
one of them is at least a P-wave state.  Furthermore, the results in
the toy model show that the P-wave state is narrower and closer to the
threshold than the S-wave state, which has analogy to the narrower
$P_c(4450)$ and wider $P_c(4380)$. Hence,  the $P_c(4450)$ and
$P_c(4380)$ may be a $5/2^+$ (P- and F-waves) state and a $3/2^-$ (S-
and D-waves) state from the $\bar{D}^*\Sigma_c$ interaction,
respectively. To confirm this assumption, a two-channel scattering
will be constructed as the toy model.

For the generating channel $\bar{D}^*\Sigma_c$, an explicit discussion
in the same quasipotential Bethe-Salpeter equation approach as
the current work has been given in Ref.~\cite{He:2015cea} with
pseudoscalar ($\pi,\ \eta$), vector ($\rho,\ \omega$) and scalar
($\sigma$) meson exchanges  included.  These two pentaquarks were
observed in the $J/\psi p$ invariant mass spectrum, so we choose it as
the observation channel. Since the $J/\psi p$ interaction is OZI
suppressed, here we  assume a potential $i{\cal V}_{J/\psi\phi\to
J/\psi\phi}=0$ as in the toy model.

The transition between generating and  observation channel is
described by  $D$ and $D^*$ exchanges. So we need the following
Lagrangians~\cite{Dong:2009tg,Lu:2016nnt,Shen:2016tzq,Garzon:2015zva}:
\begin{eqnarray}\label{eq:lag-p-exch}
{\cal L}_{\Sigma_cND^*}&=&g_{\Sigma_cND^*}\bar{N}\gamma_\mu{\bm \tau}\cdot{\bm \Sigma}_cD^{*\mu}+{\rm H.c.},\nonumber\\
{\cal L}_{\Sigma_cND}&=&-ig_{\Sigma_cND}\bar{N}\gamma_5{\bm \tau}\cdot{\bm \Sigma}_cD+{\rm H.c.}
\end{eqnarray}
Based on SU(4) symmetry,
the coupling constants  $g_{\Sigma_cND^*}$ =
3.0, and $g_{\Sigma_cND}$= 2.69~\cite{Dong:2009tg,Lu:2016nnt,Shen:2016tzq}.
The coupling of heavy-light charmed mesons to $J/\psi$ is of form ~\cite{Cheng:1992xi,Yan:1992gz,Wise:1992hn,Casalbuoni:1996pg}
\begin{eqnarray}
{\cal L}_{D^*\bar{D}J/\psi}&=&
g_{D^*\bar{D}\psi} \,  \, \epsilon_{\beta \mu \alpha \tau}
\partial^\beta \psi^\mu (\bar{D}
\overleftrightarrow{\partial}^\tau D^{* \alpha}+\bar{D}^{* \alpha}
\overleftrightarrow{\partial}^\tau D) \label{matrix3} \nonumber \\
	{\cal L}_{D^*\bar{D}^*J/\psi}&=&-ig_{D^*\bar{D}^*\psi}
\big[\psi^\mu \bar{D}^*_\mu\overleftrightarrow{\partial}^\nu D^*_\nu  -
\psi^\mu \bar D^{*\nu}\overleftrightarrow{\partial}_\mu {D}^*_\nu \nonumber\\&+&
\psi^\mu \bar{D}^{*\nu}\overleftrightarrow{\partial}_\nu D^{*}_\mu ) \big].
\end{eqnarray}
The two couplings are related to a single parameter $g_2$ as
$
g_{D^*D^*\psi} = 2 g_2\sqrt{m_\psi} m_{D^*_{(s)}}, \quad
g_{D^*D\psi}= 2 g_2 \sqrt{m_Dm_{D^*}/m_\psi }$
with $g_2=\sqrt{m_\psi}/({2m_Df_\psi})$ and $f_\psi=405$ MeV~\cite{Cheng:1992xi,Yan:1992gz,Wise:1992hn,Casalbuoni:1996pg}.
As discussed in Ref.~\cite{Lu:2016nlp}, we do not consider the form factors for the light meson coupling with $D^*$ and the $D^{(*)}$ 
coupling with $J/\psi$.  A form factor as $f(q^2)=[\Lambda^2/(\Lambda^2-q^2)]^2$, which satisfies the quark counting rule,  
is only introduced to the vertex for the baryon with a cutoff
$\Lambda$  which is chosen the same as the cutoff in the propagator for simplification.

The potential kernel can be obtained with the Lagrangians given above as our previous work in Ref.~\cite{He:2015cea}. 
In our model, only one free parameter, the cutoff $\Lambda$, is involved, which will be determined by comparison with experiment. 
As in the toy model, the poles of two-channel scattering in $3/2^-$
and $5/2^+$ waves are searched by variation of $z$ in the complex plane to satisfy
$|1-V(z)G(z)|=0$, and presented in Fig.~\ref{Fig:
DSigmaA}.   The  $J/\psi p$ invariant mass spectrum of $\Lambda_b^0\to J/\psi K^- p$ decay  is  given approximately as \cite{Hyodo:2003jw,Aceti:2014uea}
\begin{align}
	\frac{d\sigma}{dW}&=C|{\cal M}^{J^P}_{J/\psi p}|^2\lambda^\half(W^2,m_{J/\psi}^2,m_p^2)\lambda^\half
	(\tilde{W}^2,W^2,m_{K^-}^2)/W, \label{Eq: mass}
\end{align}
with $\tilde{W}$ being total energy of the decay process, that is, the mass of $\Lambda^0_b$.
The square of the $J/\psi p$ scattering amplitude $|{\cal M}^{J^P}_{J/\psi p}|^2$ is defined and 
calculated analogously to that in the toy model in Eq.~(\ref{Eq: square amplitude}).
\begin{figure}[h!]
\begin{center}
\includegraphics[bb=180 575 430 738, clip, scale=1]{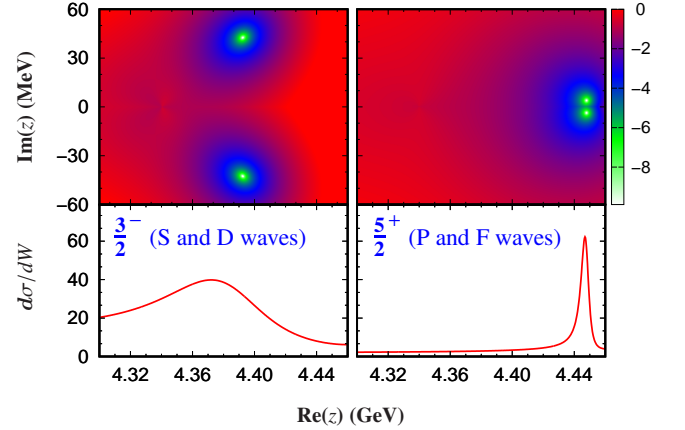}
\caption{The $\log|1-V(z)G(z)|$ and the $J/\psi p$ mass spectrum for
the $\bar{D}^*\Sigma_c$ interaction coupled with the $J/\psi p$ channel at cutoff $\Lambda$=2 GeV. The results in $\frac{3}{2}^-$ wave (left panel) and $\frac{5}{2}^+$ wave (right panel) are drawn to the same scale.
\label{Fig: DSigmaA}}
\end{center}
\end{figure}

The cutoff $\Lambda$ is varied to produce two poles which can be
related to two LHCb pentaquarks.  At cutoff $\Lambda=1.7$ GeV, which
is  close to the cutoff in the nucleon-nucleon
interaction~\cite{Gross:1991pm,VanOrden:1995eg}, a pole at $4447\pm
4i$ MeV is found in $5/2^+$-wave  $\bar{D}^*\Sigma_c$ interaction.
Correspondingly, a narrow peak appears in the $J/\psi p$ mass spectrum
near the $\bar{D}^*\Sigma_c$ threshold.  The small binding energy and
width of this state is due to the relatively weak interaction in P and
F waves. Obviously, this state can be identified as the experimentally
observed $P_c(4450)$.   In the  $3/2^-$ wave corresponding to S and D
waves, a pole at $4392+46i$ MeV is found, whhose peak is rather broad
and far from the  $\bar{D}^*\Sigma_c$ threshold because of the
relatively strong interaction in this partial wave.  Hence, as in the toy
model, the P-wave state is bound more loosely and narrower than the
S-wave state, which is consistent with the experimental observations
of the $P_c(4450)$ and the $P_c(4380)$~\cite{Aaij:2015tga}.

In Table~\ref{Tab: bound state Pc}, more results of the position of
the poles with variation of the cutoff are listed to show the
sensitivity to  the parameter, cutoff $\Lambda$. In this work we are
more interested in the bound state  in the $5/2^+$ wave near the
$\bar{D}^*\Sigma_c$  threshold.   The results show that  the mass of
this higher pole decreases by about 20 MeV with cutoff increasing from
1.6 to 1. 8 GeV.  Empirically, it is reasonable to conclude that the
result is not sensitive to the cutoff for the higher pole.  The
running of the lower pole in $\thalf^-$ wave is  faster than the
higher pole. However, considering   large experimental uncertainty of
mass  and width of this state, the result is not so sensitive to the
cutoff.

\renewcommand\tabcolsep{0.175cm}
\renewcommand{\arraystretch}{1.65}
\begin{table}[h!]
\begin{center}
\caption{The position of the poles from the $\bar{D}^*\Sigma_c-J/\psi p$ interaction with the variation of  cutoff $\Lambda$. 
The  higher and lower lines are for  $J^P=\fhalf^+$ and $\thalf^-$
waves, respectively. The  cutoff $\Lambda$ and  position  $z$ are in units of GeV and MeV, respectively. \label{Tab: bound state Pc}
}
	\begin{tabular}{c|lllll}\bottomrule[1.5pt]
		$\Lambda$ & \multicolumn{1}{c}{1.60}
		&\multicolumn{1}{c}{1.65}
&\multicolumn{1}{c}{1.70} &\multicolumn{1}{c}{1.75}
&\multicolumn{1}{c}{1.80}   \\\hline
$\fhalf^+$ & 4456+i2  & 4451+i3& 4447+i4&  4443+i5&4438+i6\\
$\thalf^-$  & 4441+i29  & 4415+i38 & 4392+i46 & 4370+i47 & 4350+i48 \\
\toprule[1.5pt]
\end{tabular}
\end{center}

\end{table}

\begin{figure}[h!]
\begin{center}
\includegraphics[bb= 180 588 460 730,clip, scale=0.98]{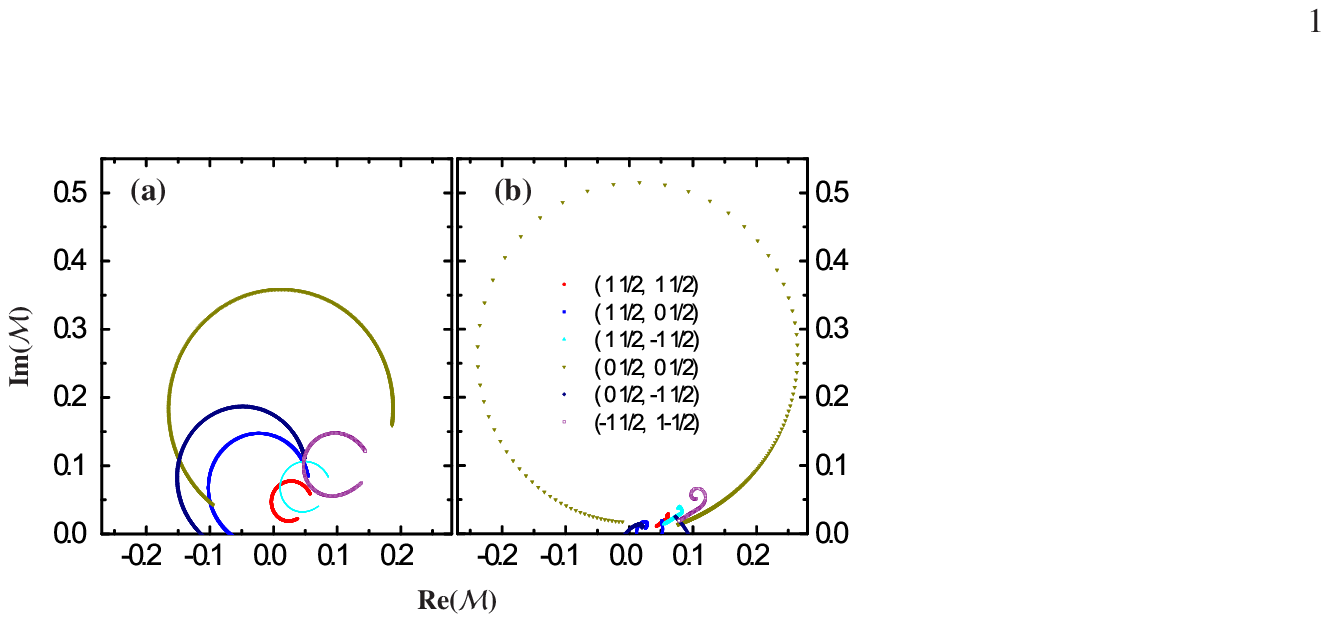}
\caption{(Color online) The Argand diagrams for the $\frac{3}{2}^-$ wave (panel a) and $\frac{5}{2}^+$ wave (panel   b). The numbers in the legend are for the helicities $(\lambda_1\lambda_2,\lambda'_1\lambda'_2)$ of the helicity amplitudes.\label{Fig: Argrand}}
\end{center}
\end{figure}

It is well known that a resonance leads to a rapid rotation of evolution of complex amplitude in the Argand diagram.
The molecular state will show a circular trajectory near the position of the peak. In Fig.~\ref{Fig: Argrand}, 
the Argand diagram are shown with the energy region from 4.30 to 4.46
GeV as in Fig.~\ref{Fig:
DSigmaA}. The energy region is divided into 500 parts, each of which is plotted as a dot in Fig.~\ref{Fig: Argrand}.

All independent helicity amplitudes are presented and it is found that
${\cal M}_{(01/2,01/2)}$ is the most important one. Because  the
amplitudes for the $J/\psi p$ scattering are adopted here,  the
results cannot be compared  directly with the LHCb
experiment~\cite{Aaij:2015tga}. But there is still something
interesting observed. The molecular states really give a rapid
rotation at energies near peaks as expected which is a behavior
characteristic of a real resonance structure. For the $5/2^+$ state
corresponding to $P_c(4450)$, almost an entire circle is formed while
for the $3/2^-$ state corresponding to $P_c(4380)$ only an arc is
formed.  The analogous behaviors can be found in the Argand diagram of
the LHCb experiment~\cite{Aaij:2015tga}. Almost all dots for the
$3/2^+$ state are at the circle, which reflects the resonance
structure cover all energy region from 4.30 to 4.46 GeV as shown in
Fig.~\ref{Fig: DSigmaA}.  But for the $5/2^+$ state, only a few dots
are involved in the rapid rotation, which reflects the small width of
this state. The dots out of the resonance region assembles  at the (0, 0)
point, and do not show a circular trajectory.

\section{Application to the $Y(4274)$}\label{sec4}

In Refs.~\cite{Liu:2010hf,He:2013oma}, the $D_sD_{s0}(2317)$
interaction has been studied and the $Y(4274)$ is assigned as an S-wave
$D_sD_{s0}(2317)$ molcular state with quantum number $J^{PC}=0^{-+}$,
which conflicts with recent LHCb
experiments~\cite{Aaij:2016iza,Aaij:2016nsc}. To reproduce the LHCb
spin parity of the $Y(4274)$ we need to introduce the P-wave
interaction. Considering that  the  $D_s\bar{D}_{s0}(2317)$
interaction is mediated by $\phi$ and $\eta$ exchanges, it is possibly
strong enough to generate a P-wave  bound state.

As in Ref.~\cite{He:2013oma}, the potential of the  $D_s\bar{D}_{s0}(2317)$  interaction by light meson exchanges can be obtained with the Lagrangian from the heavy quark field theory~\cite{Casalbuoni:1996pg},
\begin{align}
 \mathcal{L}&=
  i\frac{-2 h}{\sqrt{6}f_\pi}(D_{s}^\dag\overleftrightarrow{\partial}^\mu D_{s0}
   + D^{\dag}_{s0}\overleftrightarrow{\partial}^\mu D_{s})\partial_\mu{}\eta\nonumber\\
&-i\frac{\beta{}g_V}{\sqrt{2}}
  D_{s}^\dag\overleftrightarrow{\partial}^\mu D_{s}\phi_\mu
  +  i\frac{\beta'g_V}{\sqrt{2}}D^{\dag}_{s0}\overleftrightarrow{\partial}^\mu D_{s0}
  \phi_\mu,\label{lagrangian}
\end{align}
where the coupling constants
$h=-0.56\pm0.28$,
$\beta\beta'=0.90$, $g_V=m_\rho/f_\pi=5.8$ with $f_\pi=132$
MeV~\cite{Liu:2010hf,He:2011ed,Casalbuoni:1996pg,Isola:2003fh}.
Since $\beta$ and $\beta'$ are not well determined in the literature,
we choose $\beta\beta'=0.9\eta_{\beta\beta'}$ and
take $\eta_{\beta\beta'}$  as a free parameter.

Here we will use these potentials to study the  $1^{++}$ (P-wave)
bound state as well as the $0^{-+}$ (S-wave) bound state. To compare the contributions of S-wave and P-wave states in the $J/\psi\phi$ channel, we also introduce the transition between $D_s\bar{D}_{s0}(2317)$  and $J/\psi \phi$ channel through $D^*_{s}$ exchange. Different from the toy model and the case of $P_c(4450)$ and  $P_c(4380)$, the  $1^{++}$ (P-wave) and $0^{-+}$ (S-wave) bound states from the $D_s\bar{D}_{s0}$ interaction decay into $J/\psi\phi$ in S and P waves, respectively. The Lagrangians for $J/\psi$ coupling to $D^*_s\bar{D}_s$ and $D^*_s\bar{D}_{s0}(2317)$ reads~\cite{Casalbuoni:1996pg,Colangelo:2003sa}
\begin{align}
{\cal L}_{D^*_s\bar{D}_sJ/\psi}&=
2 g_2\sqrt{m_{D_s}m_{D^*_s}\over m_\psi}  ~\epsilon_{\beta \mu \alpha \tau}
\partial^\beta \psi^\mu \nonumber\\&~\cdot~(\bar{D}_s
\overleftrightarrow{\partial}^\tau D^{* \alpha}_s+\bar{D}^{* \alpha}_s
\overleftrightarrow{\partial}^\tau D_s), \label{matrix3} \nonumber \\
	{\cal L}_{D^*_s\bar{D}_{s0}J/\psi}&=-2g_3\sqrt{m_{D_s}m_{D_{s0}}m_\psi}
~\psi\cdot \bar{D}^*_s~D_{s0}+{\rm H.c.},
\end{align}
where $g_3={\sqrt {m_\psi}/f_\psi}$.
The Lagrangians for $J/\psi$ coupling to $D^*_s\bar{D}_s$ and $D^*_s\bar{D}_{s0}(2317)$ reads~\cite{Casalbuoni:1996pg,Colangelo:2003sa},
\begin{align}
	{\cal L}_{D^*_s{D}_s\phi}&= -i\sqrt{2}\lambda{}g_V\varepsilon_{\lambda\alpha\beta\mu}
 (D_s^{*\mu\dag}\overleftrightarrow{\partial}^\lambda
 D_s+D_s^{\dag}\overleftrightarrow{\partial}^\lambda D_s^{*\mu})
 \partial^\alpha \phi^\beta,\nonumber\\
{\cal L}_{D^*_s{D}_{s0}\phi} &=\sqrt{2}\varpi g_V(D^{\dag}_{s0}\overleftrightarrow{\partial}^\alpha D^{*\beta}_s -
  D^{*\beta\dag}_{s}\overleftrightarrow{\partial}^\alpha
  D_{s0})(\partial_\alpha{}\phi_\beta-\partial_\beta{}\phi_\alpha)\nonumber\\
&~-~\sqrt{2}\zeta{}g_V\sqrt{m_{D_{s0}}m_{D^*_s}}(D^{\dag}_{s0} D_s^{*\mu}+
  D^{*\mu\dag}_{s}D_{s0})\phi_\mu,
\end{align}
where $\lambda=0.56$ GeV$^{-1}$, $\zeta=0.727$ and $\varpi=0.364$. 
As in the case of the pentaquark, we do not consider the form factors
for the light meson coupling with $D_s$ while form factor are
introduced to the vertex for $D_{s0}^*$ because it is an excited
state. Since there does not exist experimental or theoretical
information about the form factor for $D^*_{s0}$ meson.  Form factor
as $f(q^2)=\Lambda^2/(\Lambda^2-q^2)$ are  introduced to the vertex
for the $D^*_{s0}$ meson with a cutoff $\Lambda$  which is chosen the same
as the cutoff in the propagator for simplification.

\begin{figure}[h!]
\begin{center}
\includegraphics[bb=180 575 430 738,clip, scale=1]{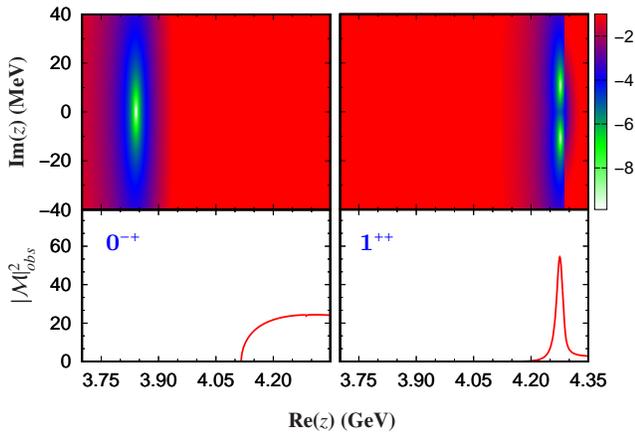}
\caption{The $\log|1-V(z)G(z)|$ and the $J/\psi \phi$ mass spectrum
for the $D_s\bar{D}_{s0}(2317)$ interaction coupled with the $J/\psi \phi$ channel at cutoff $\Lambda$=2 GeV. The results in $0^{-+}$ wave (left panel) and $1^{++}$ wave (right panel) are drawn to the same scale. The explicit partial waves on orbital angular momentum $L$ are not given here because the $1^{++}$ (P-wave) and $0^{-+}$ (S-wave) bound states from the $D_s\bar{D}_{s0}(2317)$ interaction decay into $J/\psi\phi$ in S and P waves, respectively.  \label{Fig: Y}}
\end{center}
\end{figure}

It is found that with $\Lambda=1.8$ GeV and $\eta_{\beta\beta'}=1.8$,
a pole at $4275+11i$  MeV is produced from the
$D_s\bar{D}_{s0}(2317)$ interaction  with $1^{++}$, which is presented
in Fig.~\ref{Fig: Y}.  As in the case of the toy model and the case of
the $P_c(4450)$,  the $1^{++}$ (P-wave) state appears near the threshold
while a $0^{-+}$ (S-wave) state is  far from the threshold of the
generating channel.  The pole near the  threshold at $4275\pm11i$ MeV
can be related to the Y(4274) with the spin parity  $1^{++}$
suggested by LHCb.  It is interesting to find that the $0^{-+}$
(S-wave)   state is below the $J/\psi$ threshold, which explains why
it cannot be observed in experiment.

In Table~\ref{Tab: bound state Y}, more results about the position of  the poles  are listed as in the hidden-charmed pentaquark case.
Analogously, the pole  near the $D_s\bar{D}_{s0}(2317)$ threshold  is not sensitive to cutoff. 
The mass decreases by about 10 MeV with cutoff increasing from 1.7 to 1. 9 GeV.  
The running of the lower pole in the $\thalf^-$ wave is  faster than the higher pole.
\renewcommand\tabcolsep{0.165cm}
\renewcommand{\arraystretch}{1.65}
\begin{table}[h!]
\begin{center}
\caption{The position of  poles from the $D_s\bar{D}_{s0}(2317)-J/\psi \phi$ interaction with the variation of  cutoff $\Lambda$ . The higher and lower lines are for  $J^P=1^{++}$ and $0^{-+}$ waves, respectively. The  cutoff $\Lambda$ and  position  $z$ are in units of GeV and MeV, respectively.  \label{Tab: bound state Y}
\label{diagrams}}
	\begin{tabular}{c|lllll}\bottomrule[1.5pt]
		$\Lambda$ & \multicolumn{1}{c}{1.70} &\multicolumn{1}{c}{1.75}
&\multicolumn{1}{c}{1.80} &\multicolumn{1}{c}{1.85}
&\multicolumn{1}{c}{1.90}   \\\hline
$1^{++}$  & 4278+i10  & 4276+i10 & 4275+i11 & 4272+i12 & 4269+i13 \\
$0^{-+}$ & 3900  & 3872& 3841&  3809&3774 \\
\toprule[1.5pt]
\end{tabular}
\end{center}

\end{table}

\section{Discussion and conclusion}\label{sec5}

In this work, we study the  $\bar{D}^*\Sigma_c$ and
$D_s\bar{D}_{s0}(2317)$ interactions and their relation to the
experiment observed $P_c(4450)$ and $Y(4274)$ in the hadronic molecular
state picture. The spin parities of these two states cannot be
reproduced from only S-wave interactions, so the spin parties which
correspond to P wave are considered in this work. A pole near the
$\bar{D}^*\Sigma_c$ threshold and  a pole near $D_s\bar{D}_{s0}$
threshold can be found with quantum number $5/2^+$ and $1^{++}$,
respectively. These two poles can be related to the experimentally
observed $P_c(4450)$  and $Y(4274)$.

The bound states with spin parties which correspond to S wave are also
produced as expected. When the P-wave state is produced near
threshold, the S-wave state is far from the threshold.  For the
$\bar{D}^*\Sigma_c$ interaction, the pole from  $3/2^-$-wave
interaction locates  at about 4390 MeV, which can be related to the
$P_c(4380)$ state.  As suggested in Ref.~\cite{Chen:2016qju} existence
of two or more resonant signals around 4380 MeV, especially those with
spin parity $3/2^-$, cannot be excluded because of the large widths
for the $P_c(4380)$ obtained here and in experiment. For  the
$D_s\bar{D}_{s0}(2317)$ interaction, the S-wave state is far from the
threshold even below the $J/\psi\phi$ threshold, so cannot be
observed in experiment.

By introducing an observation channel,  the effects of states in
different partial waves on experiment observables are discussed. For
the toy model and the $\bar{D}^*\Sigma_c$-$J/\psi p$ interaction, the
P-wave state near threshold is narrower than the   S-wave state far
from the threshold, but the height of the peak of the former is of the
same order of magnitude as the peak of the latter. In this work, only
two channels are included. If the width of the P-wave state is really so
smaller than S-wave state after  all possible channels included, the P-wave
state should be easy to observe in experiment.  Back to the questions
in the Introduction, at least for the cases considered in this work,
\begin{itemize}
  \item P-wave interaction is weaker but may be still enough to form a bound state.
  \item The P-wave bound state can be observed as these from S-wave interaction.
  \item The S-wave bound state should be far from the threshold if the
	  observed state corresponds tothe  P-wave bound state.
\end{itemize}

\begin{acknowledgments} This project is partially supported by the National Natural Science
Foundation of China (Grants No. 11275235 and No.11675228), and the Major State
Basic Research Development Program in China (No. 2014CB845405).

\end{acknowledgments}

\end{document}